\documentclass[12pt,oneside]{article}       %
\usepackage{amsmath, amssymb, dsfont, epsfig}
\usepackage{url}
\usepackage[usenames,dvipsnames]{color}
\usepackage[english]{babel}
\usepackage{caption}
\usepackage[comma, sort&compress]{natbib}
\usepackage{graphics}

\newcommand{\fp}[2]{\frac{\partial #1}{\partial #2}}
\newcommand{\sop}[2]{\frac{\partial^2 #1}{\partial #2^2}}
\def\EQ{{\mathbb E}_{\mathbb Q}}
\def\E{{\mathbb E}}

\begin{document}

\title{New solvable stochastic volatility models for pricing volatility derivatives
\thanks{I thank Peter Carr, Mark Craddock, Alexander Antonov, Igor Halperin, Roza Galeeva and Michael Spector for useful comments and discussion. I assume full responsibility for any remaining errors.}}
\author{Andrey Itkin \\ \\
\small Department of Finance and Risk Engineering, \\
\small Polytechnic Institute of New York University, \\
\small 6 Metro Tech Center, RH 517E, Brooklyn NY 11201, USA \\
\small email: aitkin@poly.edu \\ \\
\small Submitted to Review of Derivatives Research}

\date{}
\maketitle

\begin{abstract}
Classical solvable stochastic volatility models (SVM) use a CEV process for instantaneous variance where the CEV parameter $\gamma$ takes just few values: 0 - the Ornstein-Uhlenbeck process, 1/2 - the Heston (or square root) process, 1- GARCH, and 3/2 - the 3/2 model. Some other models were discovered in \cite{Labordere2009} by making connection between stochastic volatility and solvable diffusion processes in quantum mechanics. In particular, he used to build a bridge between solvable (super)potentials (the Natanzon (super)potentials, which allow reduction of a Schr\"{o}dinger equation to a Gauss confluent hypergeometric equation) and existing SVM. In this paper we discuss another approach to extend the class of solvable SVM in terms of hypergeometric functions. Thus obtained new models could be useful for pricing volatility derivatives (variance and volatility swaps, moment swaps).
\end{abstract}

\section{Introduction}
Analytical tractability of any financial model is an important feature. Existence of a closed-form solution definitely helps in pricing financial instruments and calibrating the model to market data. It also helps to verify the model assumptions, check its asymptotic behavior and  explain causality. In mathematical finance many models were proposed, first based on their tractability, and only then by making another argument.

Stochastic volatility models (SVM) were introduced to resolve shortcomings of the Black-Scholes model. They are highly used to evaluate derivative securities, such as equity and FX options, and variance/volatility products such as variance/volatility swaps. Tractability of these models is limited to some partial cases (which for this reason became very popular and play the same role for SVM as the Black-Scholes model for local volatility models). Classical solvable SVM use a CEV process for the instantaneous variance where the CEV parameter $\gamma$ takes just few values: 0 - the Ornstein-Uhlenbeck process, 1/2 - the Heston (or square root) process, 1- GARCH or the geometric model, and 3/2 - the 3/2 model. Closed-form solutions for the option prices written on the underlying spot were provided by using Fourier inversion if characteristic function
$\E[e^{i u X_t}]$ of the underlying process $X_t$ is known in closed form. The latter is trivial to find for $\gamma = 0$, and for $\gamma = 1/2$ it was given in \cite{Heston:93}. Later the cases $\gamma = 1/2,1,3/2$ were investigated by \cite{Lewis:2000} who for this purpose developed a method of the generalized Fourier transform. Using that Lewis derived a nice representation for the characteristic functions in the above cases, while at $\gamma = 1$ it is quite complicated and expressed via series of the Pochhammer indices. This expression could be further simplified if there is no correlation between the spot and variance Brownian motions.

For pricing variance and volatility derivatives such as variance swaps and options on them one has to know either a characteristic function of the process, or a Laplace transform of the quadratic variation of the process $\E[e^{-\lambda [X_t]}]$. For SVM $[X]_t = \int_0^t v_t dt$, where $v_t$ is the instantaneous variance. For instance, under the Heston model a fair price of the variance swaps was widely reported in the literature (see, for instance, \cite{Swishchuk2004}), while for the 3/2 model it was obtained in \cite{CarrSun} (see also \cite{ItkinCarr2010}). Carr and Sun also derived a closed form expression for the joint Fourier-Laplace transform for the 3/2 model.

Some other volatility derivatives could be priced if the characteristic function is known in closed form, for instance gamma swaps (see \cite{Lee2008}), but not the volatility swaps (for those only some approximations are available, see \cite{Swishchuk2004, Gatheral2005} ) and high moment swaps (\cite{Schoutens2005}).  Same is true if the Laplace transform of the quadratic variation is known (\cite{CarrSun}). For pricing options on variance it is sufficient to have the characteristic function of the underlying process, and then use FFT (\cite{Lee2004, Sepp2008}).

The following observation is important for calibrating term structure of, e.g. the variance swaps to market data using the above models. Consider the Heston model that uses a CIR process for the instantaneous variance which is linear in drift. This results in the fair price of the variance swap to be independent of how the volatility of variance is specified. In contrast, when the drift of the instantaneous variance process is nonlinear (e.g. quadratic as in the 3/2 model) the variance swap price does depend on volatility of variance as well as on correlation between the spot and the instantaneous variance. This provides better flexibility of the model at calibration, and makes it richer in terms of modeling variance swaps term structure.

In addition to the above mentioned solvable SVM some other models were discovered later in \cite{Labordere2009} by making connection between stochastic volatility and solvable diffusion processes in quantum mechanics. In particular, he used to build a bridge between solvable (super)potentials (the Natanzon (super)potentials, which allow reduction of the Schr\"{o}dinger equation to the Gauss confluent hypergeometric equation) and the existing SVM. Two-dimensional Kolmogorov equation has been converted into the Schr\"{o}dinger equation with a scalar potential, and it was shown that the Heston model, the geometric Brownian model (GARCH) and the 3/2 model belong to the Natanzon class. \cite{Labordere2009} also claims a new result that the case $\gamma=2$ (more rigorously volatility of volatility $\sigma(a) =\alpha +\eta a^2, \ \eta = const)$ can be solved in closed form in terms of hypergeometric functions, but to the  best of our knowledge the result has not been presented.

Despite a high popularity of these models, in the first place because of their tractability, in general they are not able to explain many empirical observations.  \cite{CarrSun} give a short survey of the existing literature which provides an empirical support, that being extracted from the market data, $\gamma > 1$.
\cite{CarrSun} consider two groups of papers: one which deals with statistical processes, and the other one
which considers statistical and risk-neutral processes. In the first group using affine drift \cite{IshidaEngle2002} estimate the $\gamma$ to be 1.71 for S\&P500 daily returns measured over a 30-year period. \cite{Javaheri2004} also estimates this process on the time series of S\&P500 daily returns, but with the CEV power constrained to either be 0.5, 1.0, or 1.5. He concludes that a power of 1.5 outperforms the other two possible choices. \cite{ChackoViceira1999} use spectral GMM estimation on the same process as in Ishida and Engle. Using the CRSP value weighted portfolio, they estimate the CEV power at 1.10 using weekly data over a 35-year period, and at 1.65 using monthly data over a 71-year period. In the second group
\cite{Jones2003} examines daily S\&P100 returns and implied volatilities over a 14-year period. Using this data with the statistical version of the affine drift CEV process, and for S\&P100 daily returns, he finds the CEV power to be 1.33.  For shorter maturity options  he concludes that jumps may be needed.
\cite{Bakshi2004} look at time series of S\&P100 implied volatilities as captured by VIX.  They find that a linear drift model is rejected in favor of a nonlinear drift model, and $\gamma = 1.27$.  They conclude that $\gamma > 1$ is needed to match the time series properties of the VIX index with the CEV models.

The above results tell us that $\gamma$ is not a constant and varies depending on the market data, time to expiration etc. So it would be desirable  to have $\gamma$ to be a parameter of the model. Also a nonlinear drift often better fits the market data than its liner counterpart. Finally, it would be feasible to have the drift being a function of time to calibrate the model to the existing term-structure of the market data (like in short-term interest rates models).

Unfortunately, we do not believe one can obey all these requirements and have a tractable model at the same time. Therefore, we need a compromise. In this paper we partly sacrifice by tractability in a sense that
the characteristic function or the Laplace transform of the quadratic variation is not required to be known in closed form. However, marginal density of the instantaneous variance process should be available in closed form. If this is the case, after the model is calibrated, prices of all volatility derivatives (variance swaps, volatility swaps, higher moment swaps, VIX futures, etc) can be computed analytically (or semi-analytically, i.e. in quadratures).

To calibrate such a model, e.g. to vanilla option prices, two approaches seem to be reasonable. The first one deals with the numerical pricing. To solve a corresponding pricing PDE finite differences together with a very efficient splitting method (e.g. \cite{HoutWelfert2007}) could be used. The total complexity of the method is $O(N\mbox{x}M) \ N$ - the number of mesh points in the spot space, $M$ - the number of mesh points in the variance space.

The second approach is almost analytical and makes use of the mixing theorem. Conditional on the whole path of the instantaneous variance from $v_0$ to $v_T$ the option price $C_i$ then is given by the Black-Scholes formula with an efficient spot price $S^{e}$ and an efficient volatility $\sigma^{e}$ (see \cite{RomanoTouzi:97}). This method is very fast due its almost analytical nature, and we recommend it for calibration of the proposed models.

The rest of the paper is organized as follows. In the next section we specify a general SVM which fits the above requirements and derive a backward Kolmogorov PDE for the density of the instantaneous variance. To solve it in Section 3 we use a Lie symmetries method to find all Lie symmetries and groups of transformations of the Kolmogorov PDE. Using the ideas of \cite{Craddock2009} we show that the density which solves the Kolmogorov equation could be represented as an inverse Laplace transform of a known function. In Section 4 we show under which assumptions on time-dependent coefficients of the model this inverse Laplace transform could be calculated analytically. We also derive closed form solutions for the volatility derivatives by computing the corresponding expectations. Next Section describes various approaches to the model calibration and presents results of some numerical tests. The last Section concludes.

The proposed model and closed form solutions for the volatility derivatives are new and constitute the main contribution of this paper. 

It is important to note\footnote{This was kindly noticed by P. Carr} that the model derived in this paper to describe dynamics of the instantaneous variance looks similar to the model which was proposed in \cite{CarrLinetsky2006} to describe dynamics of the underlying spot process for the valuation of corporate liabilities, credit derivatives, and equity derivatives. The differences are as follows: \cite{CarrLinetsky2006} use a different method (theory of scaled Bessel processes) in order to compute some moments (via expectations) of the underlying process. They did not provide a density function in the closed form. The main area of applications of their results are credit and equity derivatives. In our paper the main area are volatility derivatives and options on them. We use a different mathematical approach to find the analytical representation for the density of the instantaneous variance process and all moments, including higher moments. As a consequence, we are able to price in the closed form all the above mentioned volatility products.

\section{Specification of the SVM}
We want to analyze a rather general form of the SVM which allows pricing of path-independent contingent claims such as volatility and variance swaps and options on them. Consider a frictionless market for the underlying spot price $S_t, t\in [0,T]$ for maturity $T$. Under no arbitrage there exists a risk-neutral measure $\mathbb{Q}$ such that the prices of all non-dividend paying assets are martingales under this measure. The risk-neutral process for the underlying price is:
\begin{equation*}
d S_t = (r-q) S_t dt + S_t \sqrt{v_t} d W, \qquad t \in [0,T],
\end{equation*}

\noindent where $dW$ is a $\mathbb{Q}$ standard Brownian motion, $v_t$ is the instantaneous variance, $r, q $ are the interest and continuous dividend rates. Assume that the risk-neutral process for $v_t$ is given by
\begin{equation*}
dv_t = \left[q(t) Q(v_t) + s(t) S(v_t)\right] dt + G(t,v_t) d Z_t, \qquad t \in [0,T],
\end{equation*}
\noindent where $Z_t$ is a $\mathbb{Q}$ standard Brownian motion, whose increments have known constant correlation $\rho \in [-1,1]$ with increments in the $\mathbb{Q}$ standard Brownian motion $W_t$, i.e.:
\begin{equation*}
\EQ[d Z_t,  d W_t] =  \rho dt, \qquad t \in [0,T].
\end{equation*}

Here $q(t), s(t)$ are some deterministic functions of time, which assumed to be a model input, $Q(v_t), S(v_t)$ are some yet unknown functions of $v_t$. As it was discussed earlier, using $q(t),s(t)$ to be a function of time, versus to be a constant as in a standard specification of the SVM, allows a better calibration of the model. This is especially important if the model is used for calibration of the term structure of the variance swaps.

Note, that this is rather general representation of the SVM. Let us outline which particular functional form of $q(t),Q(v_t),s(t),S(v_t),G(t,v_t)$ admits a closed-form solution that was discussed in Introduction. To the best of our knowledge only a CEV process has been used in the literature for $G(t,v_t)$, i.e. $G(t,v_t) = v^\gamma, \ \gamma > 0$. Under this model the following solutions are known.
\begin{enumerate}
\item $\gamma = 1/2, q(t) = \kappa \theta(t), Q(v_t) = 1, s(t) = \kappa, S(v_t) = v_t$  - the Heston model (\cite{Heston:93}).
\item $\gamma = 1, q(t) = \kappa \theta(t), Q(v_t) = 1, s(t) = \kappa, S(v_t) = v_t$  - GARCH-type model. \cite{Lewis:2000} derived the characteristic function for this model which is rather complicated. It can be simplified, if $\rho = 0$.
\item $\gamma = 3/2, q(t) = \kappa \theta(t), Q(v_t) = 1, s(t) = \kappa, S(v_t) = v_t$  - 3/2 model. \cite{Lewis:2000} derived the characteristic function  for this model. Later \cite{CarrSun} extended this case and obtained  the closed form solution for the characteristic function and Laplace transform of the quadratic variation when $Q(v_t) = v_t, s(t) = s, S(v_t) = v_t^2$.
\item $\gamma = 2, q(t) = \kappa \theta(t), Q(v_t) = 1, s(t) = \kappa, S(v_t) = v_t$  - \cite{Labordere2009} claims that this model can be solved in closed form in terms of hypergeometric functions, but the result has not been presented.
\end{enumerate}

Let us also remind those features of the model that are desirable in our setup.
\begin{itemize}
\item We still want to use a CEV model of volatility of volatility, e.g.  $G(t,v_t) \propto v^\gamma, \ \gamma > 0$, but with $\gamma$ being a calibrated model parameter.
\item Specification of $q(t), s(t), Q(v_t), S(v_t)$ should preserve mean-reversion.
\item To have the model being suitable for calibration of the term structure we want to preserve time dependence of, at least, some coefficients, e.g. $q(t)$ or $s(t)$, or better both.
\end{itemize}

Based on these requirements to the proposed model consider the following SDE for the instantaneous volatility $x_t = \sqrt{v_t}$
\begin{equation} \label{volOrig}
d x_t = \left[q(t) x_t^a - s(t) x_t^b \right]d t + l(t) x_t^{\gamma+1}d Z_t
\end{equation}

Application of the Ito's lemma gives the corresponding SDE for $v_t$
\begin{equation} \label{varOrig}
d v_t = \left[2 q(t) v_t^{\frac{a+1}{2}} + l^2(t)v_t^{\gamma+1} - 2 s(t) v_t^{\frac{b+1}{2}}\right]d t + 2 l(t) v_t^{\frac{\gamma+2}{2}}d Z_t
\end{equation}

Here $a,b \in \mathbb{R}, \ a \ne b, $ are some values to be determined, $l(t)$ is some deterministic function of $t$. When deriving the
Eq.~(\ref{varOrig}) we assumed that $G(t,v_t) = 2 l(t)v_t^{\frac{\gamma+2}{2}}, \ \gamma > 0, \ l(t) > 0$. Based on the sign of $s(t)$ and $q(t)$ it could be that $a > b$ or $b > a$ to preserve mean reversion.

Our further goal is to find under which values of $a$ and $b$, and particular functional form of $q(t), s(t)$ and $l(t)$ the risk-neutral density of the $v_t$ could be found in closed form. This is not a trivial problem as it could seem at the first glance. Indeed, for a standard CEV process
\begin{equation*}
d S_t = \mu S_t d t + \delta S_t^{\gamma+1}d Z_t
\end{equation*}
\noindent with the drift $\mu = const$ and parameters $\delta, \gamma = const$ it is known that change  of  variable  $z_t = 1/(\delta |\gamma|)S_t^{-\gamma}$ reduces the  CEV  process  without  drift  ($\mu = 0$)  to  the  standard Bessel process of order $1/2 \gamma$ (see \cite{RevuzYor1999, DavidovLinetsky2001}). Then the continuous part of the risk-neutral density of $S_t$, conditional on $S_0 = S$, is obtained from the well known expression for transition density of the Bessel process. If $\mu \ne 0$, using  the  result of \cite{Goldenberg:1991} this CEV  process  could be obtained  from  the process without drift via a scale and time change
\begin{equation*}
S_t^\mu = e^{\mu t} S_{\tau(t)}^0, \qquad  \tau(t) = \frac{1}{2 \mu \gamma}\left(e^{2 \mu\gamma t} -1\right).
\end{equation*}

\cite{CarrLinetsky2006} further extended this approach. They model the price of the defaultable stock under an equivalent martingale measure as a time-inhomogeneous diffusion process {$S_t^\Delta, t > 0$} with state space $E^\Delta = (0, \infty) U {\Delta}$. Here if the process $S_t$ hits zero it is sent to the cemetery state $\Delta$ at the first hitting time of zero, $T_0$. Specification of the model looks similar to the Eq.~(\ref{volOrig}), namely: if $x$ is the initial value $S_0 = x > 0$, then the diffusion coefficient is $a(t)x^{\gamma+1}$, the drift is $[r(t) - q(t) + b(t) + c a^2(t)x^{2 \gamma}]x$, where $r(t), q(t), b(t), a(t)$ are some functions of $t$. \cite{CarrLinetsky2006} call this stock price process as the jump to default extended CEV process, or JDCEV. They also give a survey of predecessor's papers that considered a similar model with constant coefficients $a,b,c$.

Using a theory of scaled Bessel processes they further managed to  derive closed form valuation formulas for corporate bonds, credit default swaps, stock options, and other credit and equity derivatives.

In this paper we, however, use a different approach. Main reason is that to fulfil our program as it was stated above, we need to find the density of the underlying process (instantaneous volatility $x_t$) in closed form, that is not provided in \cite{CarrLinetsky2006}.

To finalize description of the model it is important to note that for $\gamma > 0$ according to Feller's classification the origin $v_t = 0$ is a natural boundary, and infinity is an entrance boundary (see \cite{DavidovLinetsky2001}).

\section{Backward equation for the density and Lie symmetries}
Under the SVM with no jumps various volatility derivatives can be represented as expectations under the risk neutral measure $\mathbb{Q}$.
Then, for instance, fair strike of the variance swap (or annualized total expected realized variance) conditional on the initial level $v_0 = [v_t | t=0]$ reads
\begin{equation*}
V_2 = \EQ\left[ \frac{1}{T}\int_0^T v_s ds \ | v_0 = v \right] = \frac{1}{T}\int_0^T  \EQ[ v_s \ | v_0 = v ]ds.
\end{equation*}

Similar expression gives a fair volatility swap value
\begin{equation} \label{volswap}
V_1 =  \EQ\left[\sqrt{\frac{1}{T} \int_0^T v_s ds} \left| v_0 = v \right. \right].
\end{equation}

European call option price written on the underlying realized variance (e.g. VIX options) is by definition
\begin{equation*}
C_2 = e^{-r T} \EQ \left[ \left(\frac{1}{T}\int_0^T x^2_s d s - K\right)^+ | \ x_0 = x \right],
\end{equation*}
\noindent with the standard notation $(y-K)^+ \equiv \max(y-K,0)$.

For the put option a similar expression reads
\begin{equation*}
P_2 = e^{-r T} \EQ \left[ \left(K - \frac{1}{T}\int_0^T x^2_s ds\right)^+ | \ x_0 = x \right].
\end{equation*}

Let $p(x, t : t', y)$ be a backward transition density from the state $y$ at time $t'$ to the state $x$ at time $t$. Let $\nu = t' - t$ be a backward time. A standard argument tells us that the expectation $u(x,\nu) = \E[ \Phi(x) \ | x_t = x]$ solves the following Cauchy problem
\begin{align} \label{Kol}
\fp{u(x,\nu)}{\nu} &= \left[q(\nu) x^a - s(\nu) x^b \right] \fp{u(x,\nu)}{x} + \frac{1}{2} l(\nu)^2 x^{2(\gamma+1)} \sop{u(x,\nu)}{x}, \\
u(x,0) &= \Phi(x_{t'}) \nonumber
\end{align}
This PDE is a backward Kolmogorov equation. The transition density $p(x, t : t', y)$ is also a fundamental solution of the Eq.~(\ref{Kol}). Thus, if we know the density, the expectation could be computed according to $u(x,t') = \int_0^\infty \Phi(y) p(x, 0 : t', y) dy$ where it was assumed that we price all products at $t=0$.

However, the fundamental solutions are not unique. In contrast, the probability transition density $p(x, t : t', y)$ for the process $x_t$ is unique and  obeys the additional condition $\int_0^\infty $p(x, t : t', y)$ dy = 1$. Therefore, it is not efficient to use standard methods to solve the Eq.~(\ref{Kol}) if one needs to determine the density. Indeed, suppose we use a change of variables method that reduces the Eq.~(\ref{Kol}) to, say, the heat equation. The density for the heat equation is known. However, under backward transformation to the original variables, this function will not integrate to 1, despite it still remains to be the fundamental solution of the Eq.~(\ref{Kol}). Therefore, it is difficult to distinguish the density from the other solutions. Moreover, as shown in \cite{CraddockLennox} this requires a theory of generalized functions and distributions.

Recently \cite{Craddock2009} proposed a new method to find the density, which utilizes Lie symmetry analysis of the parabolic PDEs. What actually Craddock showed in his paper is that for the PDEs with nontrivial Lie symmetry algebras, the Lie symmetries
naturally yield Fourier and Laplace transforms of fundamental solutions. Therefore, our further goal is to derive an explicit representation of such transforms in terms of the coefficients of the PDE, and find the solutions that obey the desirable properties of our model.

\subsection{Lie symmetries}
A symmetry  of a differential equation is a transformation which maps solutions to solutions. In the 1880s Lie developed a technique for systematically determining all groups of point symmetries for systems of differential equations. By Lie's method, we look for infinitesimal symmetries of the form (see \cite{Olver1993})\footnote{In this section for a better readability we revert our notation for the backward time $\nu$ back to $t$ since this should not create any problem.}
\begin{equation} \label{vf}
\bf{v} = \xi(x, t, u)\partial_x + \tau(x, t, u)\partial_t + \phi(x, t, u)\partial_u.
\end{equation}
We need to find conditions  on $\xi, \tau, \phi$ which  guarantee  that $\bf{v}$ generates  a  symmetry  of the Eq.~(\ref{Kol}). Standard arguments show that $\tau$  can only depend on $t$,  and $\xi$ can only depend on $x$ and $t$. Further, $\phi$ must be linear in $u$, e.g. $\phi(x, t, u) = \alpha(x,t)u$. Lie's Theorem  says  that $\bf{v}$ generates  a  local  group  of  symmetries  if  and  only  if
\begin{equation} \label{pr2}
pr^2 \bf{v}\left\{\fp{u}{t} - \left[q(t) x^a - s(t) x^b \right] \fp{u}{x} - \frac{1}{2} l^2(t) x^{2(\gamma+1)} \sop{u}{x}\right\} = 0.
\end{equation}
\noindent where $pr^2 \bf{v}$ is the second prolongation of $\bf{v}$, and $u$ is the solution of the Eq.~(\ref{Kol}). The explicit prolongation formula for a vector field is given in \cite{Olver1993}.

If $\gamma \ne 1$ this leads to
\begin{align}
\phi(x,t) &= \alpha(x,t) u + \beta(x,t), \\
\xi(x,t) &= x^{\gamma+1 } l(t) \left\{\sigma(t)+\frac{x^{-\gamma } \left[2 \tau(t) l'(t)+l(t) \tau'(t)\right]}{2 \gamma l(t)^{2}}\right\}. \nonumber
\end{align}

\noindent where $\alpha(x,t), \beta(x,t)$ are some solutions of the our backward Kolmogorov equation. Choose $\beta(x,t) = 0$. Then the only nontrivial
solution for $\alpha(x,t)$ could be obtained if $a = 1, b = 2\gamma+1$, $\alpha(x,t)$ is given by the following formula
\begin{equation*}
\alpha(x,t) = c_1 + \eta(t)x^m, \quad \eta(t) = c_2 e^{\int _0^t m q(y)d y},
\end{equation*}
\noindent where $m$ is a parameter to be chosen, and $c_1,c_2$ are the integration constants, and $s(t)$ reads
\begin{equation*}
s(t)= \frac{1}{2} (m-1) l(t)^2.
\end{equation*}

Substitution of these values into the Eq.~(\ref{volOrig}) transforms it to
\begin{equation} \label{volOrig2}
d x_t = \left[q(t) x_t - \frac{1}{2}  (m-1) l(t)^2 x_t^{2\gamma+1} \right]d t + l(t) x_t^{\gamma+1}d Z_t
\end{equation}

Now a tedious algebra (that we omit here) shows that the Eq.~(\ref{pr2}) could be solved in 3 cases: $m = -2 \gamma, -\gamma, \gamma$. Accordingly, all basis triplets $\xi(x, t, u)$, $\tau(x, t, u)$, $\alpha(x, t, u)$ that produce a vector field in the Eq.~(\ref{vf})
\footnote{Remember that $\phi(x, t, u) = \alpha(x,t)u$ } and solve the Eq.~(\ref{pr2}) are given below

\subsubsection{$m = -\gamma$}
In this case the explicit form of our model follows from the Eq.~(\ref{volOrig2})
\begin{equation} \label{model1}
d x_t = \left[\frac{1}{2}  (\gamma+1) l(t)^2 x_t^{2\gamma+1} + q(t) x_t\right]d t + l(t) x_t^{\gamma+1}d Z_t.
\end{equation}

Lie algebra of infinitesimal symmetries of the Eq.~(\ref{Kol}) is now spanned by the vector fields
\begin{align} \label{sym}
{\bf v}_1 &= u \partial_u \\
{\bf v}_2 &= \gamma Z(t) \left(\int_0^t \frac{l(y)^2}{Z(t)^2} dy \right) x^{1+\gamma } \partial_x +  \frac{1}{Z(t)} x^{-\gamma} u \partial_u \nonumber \\
{\bf v}_3 &= Z^2(t) \left(\frac{l(0)}{l(t)} \right)^2 \left[\partial_t - q(t) x \partial_x \right] \nonumber \\
{\bf v}_4 &= - x \left[\frac{1}{2\gamma} + q(t) \kappa(t) \right] \partial_x
+ \kappa(t) \partial_t, \qquad \kappa(t) = \left(\frac{Z(t)}{l(t)}\right)^2\int_0^t \left(\frac{l(y)}{Z(y)}\right)^2 dy \nonumber \\
{\bf v}_5 &= Z(t) l(0) x^{1+\gamma} \partial_x \nonumber
\end{align}
Here $Z(t) \equiv e^{\int_0^t \gamma  q(y) dy}$. Among these solutions we are interested just in non-trivial ones with $\alpha(x,t) \ne 0$ and $\xi(x,t) \ne 0$, i.e. those in line 2 of the Eq.~(\ref{sym}).

According to the Lie's method (see Chapter 2 of \cite{Olver1993}) given the symmetry $\bf{v}_i$ in the form of the Eq.~(\ref{vf}) we need to exponentiate it in order to find the one-parameter group $G_i$ generated by the $\bf{v}_i$.

The first symmetry in the Eq.~(\ref{sym}) is trivial and says that the new solution of the Eq.~(\ref{Kol}) can be obtained from another solution by multiplying it by a constant. The symmetries in lines 3,4 are hard to reproduce in closed form, since exponentiation can not be done for the arbitrary time-dependent functions $q(t)$ and $l(t)$\footnote{At the best the result can be expressed via inverse functions}. The symmetry in line 5 translates just $x$, so
\begin{equation*}
G_5: \left\{
\left[x^{-\gamma}+\gamma  \mu Z(t) l(0)\right]^{-1/\gamma}, t, u \right\},
\end{equation*}

Exponentiating line 2 of the Eq.~(\ref{sym}) we obtain
\begin{equation} \label{Gmap}
G_2: \left\{
\left[x^{-\gamma}+\gamma  \mu \xi_c(t)\right]^{-1/\gamma}, t, \exp\left[-\frac{x^{-\gamma} \mu + \frac{1}{2}\gamma \mu^2 \xi_c(t)}{Z(t)}\right] u \right\},
\end{equation}
\noindent where $\xi_c(t) = \gamma Z(t) \left(\int_0^t Z(y)^{-2} l(y)^2 dy\right)$, $u$ is some solution of the Eq.~(\ref{Kol}), and $\mu$ is the group parameter.

Existence of this symmetry group implies that if $u = G(x, t)$ is a solution of the Eq.~(\ref{Kol}) so are the functions
\begin{equation*}
u_\mu(x,t) = \exp\left\{- x^{-\gamma } \mu  Z(t)^{-1} + \frac{1}{2} \gamma Z(t)^{-1} \xi_c(t) \mu^2 \right\}
G\left(\left[x^{-\gamma }+\gamma  \mu  \xi_c(t)\right]^{-1/\gamma },t\right).
\end{equation*}
Note, that the corresponding SDE for the instantaneous variance can be derived from the Eq.~(\ref{model1}) using Ito's lemma
\begin{equation*}
d v_t = \left[(2+\gamma) l(t)^2 v_t^{1+\gamma } + 2 q(t) v_t \right]d t + 2 l(t) v_t^{\frac{\gamma+2}{2}}d Z_t.
\end{equation*}
The first term in the drift is positive for if $\gamma > -2$, therefore to preserve mean reversion we must have $q(t) < 0$. Also $v_t^{1+\gamma }$ grows faster than $v_t$ if $\gamma > 0$, therefore the ratio $q(t)/l(t)^2$ has to be a rapidly increasing function of time to compensate this effect and provide a mean reverting behavior of the drift. Finally, based on our earlier discussion of the empirical data we expect the CEV exponent to vary from 1 to 2, which means $0 < \gamma < 2$.

\paragraph{Specific form of $q(t)$ and $l(t)$.} Despite we are not able to analytically exponentiate all symmetries found in above, we can extend the class of tractable symmetries by choosing some particular form of $q(t)$ or $l(t)$. Assume that
\begin{equation} \label{qf}
q(t) = \frac{1}{\gamma}\frac{l'(t)}{l(t)}.
\end{equation}

Under this assumption our models takes the form
\begin{equation} \label{volOrigPart}
d x_t = \left[\frac{1}{2} x^{1+2 \gamma } (1+\gamma) l(t)^2+\frac{x l'(t)}{\gamma  l(t)} \right]d t + l(t) x_t^{\gamma+1}d Z_t.
\end{equation}

The first term of the drift is positive. Therefore, to preserve mean reversion we need $l(t)$ to obey the condition
$l'(t)/l(t) = - \theta H(t)$ where $\theta$ is a calibration constant, and $H(t)$ is some increasing function of $t$. To obey this condition we choose
\begin{equation} \label{lf}
l(t) = \epsilon e^{-\int _0^t \theta H(y)dy},
\end{equation}
\noindent where $\epsilon$ is a constant volatility of volatility. With this expression for $l(t)$ our model for $v_t$ now reads
\begin{equation*}
d v_t = \left[(2+\gamma) l(t)^2 v_t^{1+\gamma } - 2\frac{ \theta H(t)}{\gamma } v_t \right]d t + 2 l(t) v_t^{\frac{\gamma+2}{2}}d Z_t.
\end{equation*}

Again two concurrent effects affect the drift. On the one hand $v^{\gamma+1}$ grows faster than $v$ if $\gamma > 0$ that does not allow mean-reversion. On the other hand  the exponent in the first term makes it rapidly decreasing if the function $H(t)$ grows fast with time. Thus, one can always properly choose $H(t)$ to guarantee mean reversion.

With functions $q(t), l(t)$ which obey the Eq.~(\ref{qf}),(\ref{lf}) symmetry ${\bf v_4}$ in the Eq.~(\ref{sym}) reads
\begin{equation*}
{\bf v}_4 = \frac{2\theta H(t) t -1}{2\gamma} x \partial_x + t \partial_t,
\end{equation*}

\noindent that now could be exponentiated in closed form.

Some other choices of the function $l(t)$ are also possible for this purpose, for instance
\[
l(t) = \frac{\epsilon}{t}\exp\left[-\frac{1}{2t}+ \gamma  \int_0^t q(z)dz\right],
\]
\noindent that gives ${\bf v}_4 = - x \left[\frac{1}{2\gamma} + q(t) t^2 \right] \partial_x + t^2 \partial_t$, etc.

\subsubsection{$m = \gamma$}
In this case the explicit form of our model is
\begin{equation} \label{model2}
d x_t = \left[\frac{1}{2}  (1-\gamma) l(t)^2 x_t^{2\gamma+1} + q(t) x_t\right]d t + l(t) x_t^{\gamma+1}d Z_t,
\end{equation}
\noindent and
\begin{equation*}
d v_t = \left[(2-\gamma) l(t)^2 v_t^{1+\gamma } + 2 q(t) v_t \right]d t + 2 l(t) v_t^{\frac{\gamma+2}{2}}d Z_t.
\end{equation*}
Therefore, again to preserve mean-reversion either $q(t)$ must be negative and $\gamma < 2$ whereas $l(t)/q(t)$ must be a rapidly decreasing function of $t$, or $\gamma > 2$ and $q(t) > 0$.

Lie algebra of infinitesimal symmetries of the Eq.~(\ref{Kol}) in this case is spanned by the vector fields
\begin{align*}
{\bf v}_1 &= u \partial_u \\
{\bf v}_2 &= \frac{Z(t)}{\gamma}x^{1+\gamma} \partial_x + Z(t) x^\gamma u \partial_u \nonumber \\
{\bf v}_3 &= - x \left[\frac{1}{2\gamma} + q(t) \kappa(t) \right] \partial_x
+ \kappa(t) \partial_t  \nonumber \\
{\bf v}_4 &= - Z(t)^2 \left[\frac{l(0)}{l(t)}\right]^2 q(t) x \partial_x + \kappa(t) \partial_t \nonumber
\end{align*}
Exponentiating line 2 we obtain
\begin{equation} \label{Gmap2}
G_2: \left\{
\left[x^{-\gamma}+ \mu Z(t)\right]^{-1/\gamma}, t, \left[1 + \mu Z(t) x^\gamma \right] u \right\}.
\end{equation}

\subsubsection{$m = -2 \gamma$} \label{model3SDE}
In this case the explicit form of our model is
\begin{equation} \label{model3}
d x_t = \left[\frac{1}{2}  (1+2 \gamma) l(t)^2 x_t^{2\gamma+1} + q(t) x_t\right]d t + l(t) x_t^{\gamma+1}d Z_t,
\end{equation}
\noindent and
\begin{equation} \label{model3Var}
d v_t = 2\left[(1+\gamma) l(t)^2 v_t^{1+\gamma } + q(t) v_t \right]d t + 2 l(t) v_t^{\frac{\gamma+2}{2}}d Z_t.
\end{equation}

Lie algebra of infinitesimal symmetries of the Eq.~(\ref{Kol}) here is spanned by the vector fields
\begin{align*}
{\bf v}_1 &= u \partial_u \\
{\bf v}_2 &= 2 \gamma \left[ \int_0^t \left(\frac{l(y)}{Z(y)}\right)^2 dy
+ 2 \gamma q(t) \kappa_1(t)\right] x \partial_x -4 \gamma^2 \kappa_1(t) \partial_t
+ Z(t)^{-2} x^{-2\gamma} u \partial_u \nonumber \\
& \kappa_1(t) = \frac{Z(t)^2}{l(t)^2}\left[\int_0^t \left(\frac{l(z)}{Z(z)}\right)^2\left(\int_0^z
\left(\frac{l(u)}{Z(u)}\right)^2 du \right) dz \right]  \nonumber \\
{\bf v}_3 &= - x \left[\frac{1}{2\gamma} + q(t) \kappa(t) \right] \partial_x
+ \kappa(t) \partial_t \nonumber \\
{\bf v}_4 &= Z(t)^2 \left[\frac{l(0)}{l(t)}\right]^2 \left[-q(t)x \partial_x + \partial_t \right] \nonumber
\end{align*}

Only line 2 has a particular interest for our purposes (see next section) since it provides a non-trivial transformation of the solution, not just the coordinates. Unfortunately, it can not be exponentiated at arbitrary $l(t)$. and $q(t)$. Therefore, we will use their specific form proposed in the Eq.~(\ref{qf}) and (\ref{lf}). Then the vector field $\bf v_2$ reduces to
\begin{equation*}
{\bf v}_2 = -2 t \gamma  \epsilon ^2 [-1+t \theta  H(t)] x \partial_x -2  \gamma ^2 \epsilon ^2 t^2 \partial_t
+e^{2 \int_0^t \theta  H(y) dy} x^{-2\gamma} u \partial_u
\end{equation*}

To exponentiate it, choose the simplest model $H(t) = \gamma$. Thus, in this model the speed of mean reversion is determined by the only parameter $\theta$. Exponentiating, we find some new solutions for our problem
\begin{align} \label{so8}
u_\mu(x,t)  &= \exp\left[-\frac{e^{2 t \gamma  \theta } x^{-2 \gamma } \mu }{1+2 t \gamma ^2 \mu }\right] \\
& \cdot G\left(e^{t \theta \left(-1+\frac{1}{1+2 t \gamma ^2 \mu }\right)} x \left(1+2 t \gamma ^2 \mu\right)^{1/\gamma },\frac{t}{1+2 t \gamma ^2 \mu }\right). \nonumber
\end{align}

\section{Generalized Laplace transform and solutions for the density}
 The main idea of \cite{Craddock2009} is that as the density $p(x,t : y,t')$ is also the Green's function of the backward Kolmogorov PDE, the solution of this PDE can be represented in the form \footnote{In this section for the backward time $t-t'$ we again use the notation $\nu$}
\begin{equation} \label{green}
u_\mu(x,\nu) = \int_0^\infty u_\mu(y,0) p(x,t : y,t') dy
\end{equation}

Now let us use the solutions map of the Kolmogorov PDE found in the previous section. Few useful observations could be made immediately.
\begin{enumerate}
\item $G(x,t) = 1$ is the solution of the Kolmogorov PDE given in the Eq.~(\ref{Kol}). Therefore, we plug in this solution into all the symmetry maps from the previous section.
\item By definition $Z(\nu) \equiv e^{\int_0^\nu \gamma  q(y) dy}$. Switching back to time $t'$ and $t$ translates $Z(t,t')$ to
\begin{equation*}
Z(t,t') \equiv e^{\int_t^{t'} \gamma  q(y) dy}, \quad t' \ge t.
\end{equation*}

\noindent Also by obvious reasons only positive values of the initial level of the instantaneous volatility $x$ and the volatility of volatility speed $l(t)$ are considered here. Therefore, rewriting $\xi_c(\nu)$ by switching back to time $t$ and $t'$ shows that $\xi_c(t,t')$ is positive:
\begin{equation*}
\xi_c(t,t') = 2\gamma Z(t,t') \int_t^{t'} Z(y,t')^{-2} l(y)^2 dy \ge 0.
\end{equation*}

\item At $\nu=0$ (or $t=t'$) $Z(t',t') = 1, \ \xi_c(t',t') = 0$.
\end{enumerate}

Let us start with the solutions map found for $m=-\gamma$ and given in the Eq.~(\ref{Gmap}), i.e. with the model given in the Eq.~(\ref{model1}). Substituting this map into the Eq.~(\ref{green}) and taking into account the above observations we obtain.
\begin{equation} \label{greenL}
\int_0^\infty  e^{-y^{-\gamma } \mu}  p(x,t : y,t') dy = e^{-x^{-\gamma } \mu  Z^{-1}(t,t') + w(t,t') \mu^2}
\end{equation}
\noindent where $w(t,t') = \frac{1}{2} \gamma^2 \left(\int_0^t Z(y)^{-2} l(y)^2 dy\right) > 0$.

But the lhs of this equation is just a generalized Laplace transform of the density, exactly in accordance with the main idea of the method in \cite{Craddock2009}.

Few useful theorems justifying this fact are proven in \cite{Craddock2009}, where we refer the reader to if she is interesting in more details on this method. Below a short summary of the results necessary for our further steps is presented. First, to prove that the rhs of the Eq.~(\ref{greenL}) is
a generalized Laplace transform, observe that this is correct if the rhs is a Laplace transform. This follows from a simple change of variables
$y^{-\gamma} = z$. Also $u_\mu(x,t) = e^{-x^{-\gamma } \mu  Z(t,t') + w(t,t') \mu^2}$ is analytic in $1/\mu$, and any function analytic in $1/\mu$ is automatically a Laplace transform. Thus, $u_\mu(x,t)$ is a generalized Laplace transform of some distribution.

Therefore, we can find fundamental solutions by inverting this generalized Laplace transform. Thus found fundamental solution is the transition density. To prove, let $\mu = 0$ in the Eq.~(\ref{greenL}) which gives $\int_0^\infty  p(x,t : y,t') dy = 1$.

Now to invert the transform, make change of variables $z = y^{-\gamma}$ and rewrite the Eq.~(\ref{greenL}) as
\begin{equation*}
\frac{1}{\gamma}\int_0^\infty  e^{-z \mu}  z^{-\frac{\gamma+1}{\gamma}} p(x,t : z^{-1/\gamma},t') dz = e^{-x^{-\gamma } \mu  Z(t,t') + w(t,t') \mu^2}.
\end{equation*}

Therefore,
\begin{equation} \label{densArb}
p(x,t : y,t') =  \gamma y^{-(\gamma+1)}\mathcal{L}^{-1}_{\mu,z} \left[e^{-x^{-\gamma } \mu  Z(t,t') + w(t,t') \mu^2}\right]\Big | _{z \rightarrow y^{-\gamma} }
\end{equation}

The Bromwich integral in the rhs is well defined since the integrand vanishes at $\mu \rightarrow i \infty$ and $\mu \rightarrow -i \infty$. Unfortunately, to the best of our knowledge the inverse Laplace transform in the rhs has no closed form representation.

We can run same machinery for the symmetry map found at $m = -2\gamma$ and presented in the Eq.~(\ref{so8}), i.e. for the model given in the Eq.~(\ref{model3}). This gives rise to the following representation of the transition density
\begin{equation} \label{green2}
\int_0^\infty  e^{-y^{-2\gamma } \mu}  p(x,t : y,t') dy = \exp\left[-\frac{e^{2 (t'-t) \gamma  \theta } x^{-2 \gamma } \mu }{1+2 (t'-t) \gamma ^2 \epsilon^2 \mu }\right].
\end{equation}

Inverting this Laplace transform we obtain
\begin{align} \label{dens}
p(x,t : y,t') &= 2\gamma y^{-(2 \gamma+1)} \exp\left[-\frac{e^{2 (t'-t) \gamma \theta} x^{-2 \gamma }+y^{-2 \gamma }}{2 (t'-t) \gamma ^2 \epsilon^2}\right] \nonumber \\
&\cdot \left[\frac{e^{(t'-t) \gamma \theta} x^{-\gamma } y^{\gamma } }{2 (t'-t) \gamma ^2 \epsilon^2}
 I_1\left(\frac{e^{(t'-t) \gamma \theta} x^{-\gamma } y^{-\gamma }}{(t'-t) \gamma ^2 \epsilon^2}\right) + \delta\left(y^{-2 \gamma }\right)\right],
\end{align}
\noindent where $I_d(x)$ is the modified Bessel function of order $d$.

To verify that this is the density let us check that $p(x,t : y,t')$ given by this expression integrates to 1. Omitting an intermediate algebra we find
\begin{equation*}
\int_0^\infty p(x,t : y,t') dy = 1 - \exp\left[-\frac{e^{2 (t'-t) \gamma \theta} x^{-2 \gamma }}{2 (t'-t) \gamma ^2 \epsilon^2}\right] \Psi(0) = 1,
\end{equation*}
\noindent where $\Psi(x)$ is the Heaviside theta function, and it is assumed that $\Psi(0) = 0$ since we work in a space of left-continuous functions.

On the other hand the direct substitution of $\mu=0$ into the Eq.~(\ref{green2}) verifies that $p(x,t : y,t')$ integrates to 1 unless $\gamma > 0$ and $x \ne 0$. Therefore, strictly speaking the above solution for the density can not be used for $\gamma > 0$. In the next section we will discuss this in more detail.

For the third model found at $m=\gamma$ and given in the Eq.~(\ref{model2}) we can use the solution map in the Eq.~(\ref{Gmap2}). Substitution of this map into the Eq.~(\ref{green}) gives rise to the following integral equation for the transition density
\begin{equation*}
\int_0^\infty  y^{\gamma } p(x,t : y,t') dy = Z(t) x^\gamma.
\end{equation*}
The lhs of this equation is not the Laplace transform. Therefore, we can not use the Craddock method to find thus defined transition density. However, one can treat this expression either as the Mellin transform\footnote{This was pointed out by A. Antonov}, or just directly guess that the density which solves this equation is
\begin{equation} \label{trDelta}
p(x,t : y,t') = \delta\left[y - x e^{\int_t^{t'} q(s) ds}\right]
\end{equation}

\section{Pricing variance and volatility swaps}
\subsection{Specific form of $q(t)$ and $l(t)$ given in the model Eq.~(\ref{model3}) ($m = -2\gamma$)}
Since the density of the volatility process $x_t$ is known we can price various volatility moments of the order $i = 1,2...$ by computing the corresponding expectation and then, if necessary, integrating the result in time. We can do it explicitly in the special cases discussed in Section~\ref{model3SDE}.

However, as it was mentioned already there are two problems with this solution for the transition density. The first one is that at $x \rightarrow 0$
and $\mu \rightarrow 0$ the behavior of the product $x^{-2 \gamma} \mu$ is undefined if $\gamma> 0$. To highlight the second problem let us formally assume that the first problem is resolved, and the Eq.~(\ref{dens}) defines the correct transition density at $\gamma > 0$. Then the fair price of the $i$-th moment reads
\begin{align} \label{spVi}
V_i = \frac{1}{T}\int_0^T & Q(0,t',x) dt', \\
Q(t,t',x) &= \int_0^\infty y^{i} p(x,t : y,t') dy \nonumber \\
= 2^{\frac{1-i}{2\gamma }} &\exp\left[-\frac{e^{2 (t'-t)\gamma \theta } x^{-2 \gamma }}{2 (t'-t) \gamma^2 \epsilon^2}+2 (t'-t) \gamma \theta \right]  \left[(t'-t) \gamma ^2 \epsilon^2 \right]^{\frac{1-i}{2 \gamma }} \nonumber \\
&\cdot  \Gamma\left(\frac{1-i+4 \gamma }{2 \gamma }\right) M\left(\frac{1-i+4 \gamma }{2 \gamma },2,\frac{e^{2 (t'-t) \gamma \theta } x^{-2 \gamma }}{2 (t'-t) \gamma ^2 \epsilon^2}\right) x^{-2 \gamma } \nonumber \\
= G &\phi^{1-\kappa} \Gamma(2-\kappa)M(\kappa,2, -G), \nonumber \\
& \qquad G \equiv e^{2 (t'-t)\gamma \theta } x^{-2 \gamma }/\phi, \quad, \phi \equiv 2 (t'-t) \gamma^2 \epsilon^2, \quad \kappa = \frac{i-1}{2\gamma}. \nonumber
\end{align}
\noindent Here $\Gamma(x)$ is the gamma function, $M(a,b,x)$ is the Kummer confluent hypergeometric function.

Therefore, in the proposed SVM the price, e.g. of the variance swap, can be expressed via hypergeometric functions, similar to what was discussed in Introduction, and in a more general case in \cite{BSHyper2001}. Also as shown in \cite{ItkinCarr2010} under the $3/2$ model the variance swaps price has the closed form representation also in terms of the confluent hypergeometric function.

The internal integral $Q(0,t',x)$ exists under some conditions on $i$. In particular, when $\gamma \ge 0$ it always exists at $i=1$. If $i=2$ the existence condition   requires $ \gamma > 1/2$. At $i=3$ the condition is $\gamma > 1$ and at $i=4$ the condition reads $\gamma > 3/2$. Therefore, these conditions being translated to the CEV exponent $\beta = (2+\gamma)/2$ in the Eq.~(\ref{model3Var}) mean that first four moments of the pdf for the instantaneous volatility can be computed using the above expression, if $\beta$ varies in the ranges given in Tab~\ref{ranges}.

\begin{table}[t!]
\begin{center}
\begin{tabular}{|c|l|}
\hline
$i$ & Range \cr
\hline
1 & $\beta > 1$ \cr
2 & $\beta > 1.25$ \cr
3 & $\beta > 1.5$ \cr
4 & $\beta > 1.75$ \cr
\hline
\end{tabular}
\caption{Ranges of the CEV exponent $\beta = (2+\gamma)/2$ where the Eq.~(\ref{spVi}) can be used to compute the $i$-th moment of the instantaneous volatility pdf.}
\label{ranges}
\end{center}
\end{table}

Also note, that the expression for $V_i$ in the Eq.~(\ref{spVi}) is well-behaved at $t \rightarrow t'$. Expanding the gamma and Kummer functions into series at $t \rightarrow t'$ it is possible to show that in this limit $\ V_i \rightarrow x^{i -2 \gamma -1}$.

The problem with this solution is that $y(G) = G M(\kappa,2, -G)$ is an increasing function of $G$ since $\kappa \le 0$. As $G \propto x^{-2\gamma}$ this means that $V_i$ is the decreasing function of $x$. This fact is in contradiction with a usual observation that the variance and volatility swaps price increases with the increase of the initial level $x$.

This problem can be resolved, however, if we assume that $\gamma$ is negative. Based on the Eq.~(\ref{model3Var}) suitable values of $\gamma$ lies in the range $-1 < \gamma < 0$. This has few advantages. First we can relax the conditions issued on the explicit form of $l(t)$ because with these values of $\gamma$ the variance equation Eq.~(\ref{model3Var}) is always mean-reverting as far as $q(t) < 0$. Second, this range of $\gamma$ eliminates any restrictions on the existence of the solution in the Eq.~(\ref{spVi}) for the arbitrary $i$. Third, the price now increases with the increase of the initial level $x$, thus following the observable behavior of these instruments.

It could seem, however, that a drawback of negative gammas is that under these conditions the CEV exponent varies from 1/2 to 1, which is not a favorable region (see the discussion in Introduction). On the other hand, we can not completely rely on the results in the cited papers for the following reason. In our model the volatility of volatility in the Eq.~(\ref{volOrig}) depends on both $\gamma$ and $l(t)$. This makes some problems when trying to find their values by calibration. Indeed, various pairs of $\gamma$ and $l(t)$ can produce same values of the product $l(t)x^{\gamma+1}$. This makes the calibration to be ambiguous if we calibrate the model just to the plain vanilla option. A possible resolution of that is to calibrate the model to both plain vanilla options and to variance swaps written on the same underlying (e.g. S\&P 500 and VIX). This would allow an unambiguous calibration and separation of the effect of $l(t)$ from that of $\gamma$ on the price of chosen calibration instruments. When computing prices of the variance swaps the Eq.~(\ref{spVi}) can be utilized. The results of such a calibration will be presented elsewhere.

We can also use the Eq.~(\ref{model3}) but applying it not to the instantaneous volatility, but to the instantaneous variance. This will allow the CEV exponent to vary from 0 to 1, thus extending the previous range.

To calculate volatility swaps we need to know the expectation in the Eq.~(\ref{volswap}) which can not be computed exactly given just the moments of the instantaneous volatility pdf. However, it could be found using approximation from \cite{BrockhausLong2000}
\begin{equation*}
\EQ[\sqrt{V}] \approx \sqrt{\EQ[V]} - \frac{Var[V]}{8 \EQ[V]^{3/2}}, \qquad V = \frac{1}{T}\int_0^T Q(0,t',x) dt'.
\end{equation*}

Since now we want to use the Eq.~(\ref{model3}) for the instantaneous variance, this formula translates to
\begin{equation*}
\EQ[\sqrt{V}] \approx \sqrt{V_1} - \frac{Var[V]}{8 V_1^{3/2}},
\end{equation*}
\noindent where $V_1$ is given in the Eq.~(\ref{spVi}). As $Var[V] = \EQ[V^2] - (\EQ[V])^2 = \EQ[V^2] - V_1^2$ the remaining goal is to find $\EQ[V^2]$. This is
\begin{equation*}
\EQ[V^2] = \frac{1}{T^2}\int_0^T \int_0^T Q(0,s,x)Q(0,t,x)ds dt.
\end{equation*}

\begin{figure}
\centering
\fbox{\includegraphics[totalheight=2.5in]{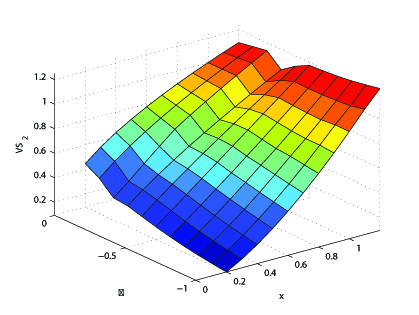}}
\caption{Price of the variance swap as the function of $\gamma$ and the initial level $x$ at $T=0.5$}
\label{VarNeg}
\end{figure}

\begin{figure}
\centering
\fbox{\includegraphics[totalheight=2.5in]{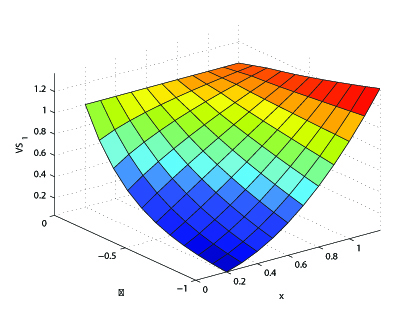}}
\caption{Price of the volatility swap as the function of $\gamma$ and the initial level $x$ at $T=0.5$}
\label{VolNeg}
\end{figure}

In Fig~\ref{VarNeg},\ref{VolNeg}  the variance swap and volatility swap prices computed using the above approach are given as a function of $\gamma$ and $x$. Here we used a model of $l(t)$ given by the Eq.~(\ref{lf}) with $H(t) = \gamma$ and $\theta = 0.1$. Also in these tests we used $\epsilon = 0.1$ and $T=0.5$. The prices are presented in volatility point divided by 100.

\begin{figure}
\centering
\fbox{\includegraphics[totalheight=2.5in]{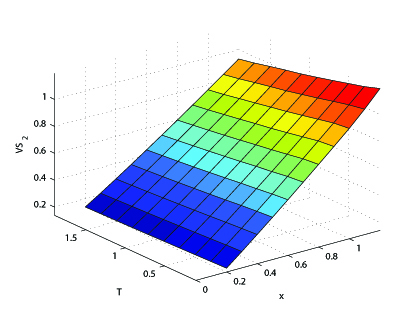}}
\caption{Price of the variance swap as the function of $T$ and the initial level $x$ at $\gamma$ = -0.6.}
\label{VS2-TNeg}
\end{figure}

In Fig~\ref{VS2-TNeg} same results are given in coordinates $(x,T)$ at $\gamma$=-0.6, $\theta$ = 0.3, $\epsilon$ = 0.1. It is seen that the variance swap price slightly decreases with the increase of $T$, that corresponds to the behavior of many SVM, where the long term run is a deterministic function of time, for instance, the Heston model.

\subsection{Solutions for $\gamma > 0$, model in the Eq.~(\ref{model2}) ($m = \gamma$)}
Since for this model the transition density found in the Eq.~(\ref{trDelta}) is just a delta function, all volatility moments could be priced in closed form, namely
\begin{equation*}
V_i = x^i \frac{1}{T}\int_0^T e^{i \int_0^{t'} q(y) dy} dt'.
\end{equation*}

Therefore, in this model the price of these derivatives does not depend on vol-of-vol $l(t)$ as well as on $\gamma$. At given $T$ the price $V_i$ is just a linear function of the $i$-th power of the initial level $x$. This is partly similar to the SVM with a linear mean-reverting drift for the instantaneous variance (like the Heston model) where the VIX price is linear in the initial level $v_0$ \cite{Lin2007}.

\subsection{General case of arbitrary $q(t)$ and $l(t)$, model in the Eq.~(\ref{model1}) ($m = -\gamma$). }

The transition density for the model Eq.~(\ref{model1}) was found as an inverse Laplace transform, which however does not have a closed form representation. Let us formally substitute this density given in the Eq.~(\ref{densArb}) into the definition of the volatility moments
\begin{equation*}
V_i = \frac{1}{T}\int_0^T \left[\int_0^\infty
\gamma y^{i - \gamma - 1} \mathcal{L}^{-1}_{\mu,z} \left[e^{-x^{-\gamma } \mu  Z(0,t') + w(0,t') \mu^2}\right]\Big | _{z \rightarrow y^{-\gamma} }
\right] dt'
\end{equation*}

Since we don't know the inverse Laplace transform under the integral in closed form it could seem that this double integral could be computed just numerically. However, there is a trick which helps to find a closed form representation of the internal integral. The idea of the trick consists in differentiating both parts of the  Eq.~(\ref{greenL}) by $\mu$ and then putting $\mu = 0$. If we do this just once the result reads
\begin{equation*}
\int_0^\infty  y^{-\gamma } p(x,t : y,t') dy = x^{-\gamma } Z(t,t')
\end{equation*}

In other words the lhs of this expression is the expected value $\EQ[x^{-\gamma} | x_0 = x]$. Since we need to compute the expectation $\EQ[x^i | x_0 = x], \quad i \in \mathbb{Z}$ the above result does not seem to be directly translated into the necessary form. Nevertheless, this could be done using a theory of fractional derivatives.

Indeed, following \cite{FracDer1976} consider Weyl fractional derivative which is defined as
\begin{equation} \label{fdd}
D^\alpha_{x,-\infty} F(x) = \frac{(-1)^{-\alpha}}{\Gamma(-\alpha)}\int_x^\infty F(t)(t-x)^{-\alpha-1}dt,
\end{equation}

\noindent The beauty of this representation consists in the fact that, e.g. for the exponential function differentiation keeps its standard rules regardless whether the derivative order is integer or real, i.e.
\begin{equation*}
D^\alpha_{x,-\infty} e^{b x} = b^\alpha e^{b x}, \qquad Re(b) < 0.
\end{equation*}

Going back to the Eq.~(\ref{greenL}) the only term in the lhs which depends on $\mu$ is $e^{-y^{-\gamma } \mu}$. Or using the notation of the previous equation, this is $e^{b \mu}, \ b = -y^{-\gamma} < 0$. Taking the fractional derivative of the order $\alpha$ of this expression we expect to obtain $c y^i e^{b \mu}$, where $c$ is some constant. That gives $b^\alpha = c y^i$ or $\alpha= -i/\gamma, c=(-1)^{-i/\gamma}$.

Further to compute this fractional derivative of the rhs of the Eq.~(\ref{greenL}) let us use the definition of $D^\alpha_{x,-\infty}$ in the Eq.~(\ref{fdd}). Also after taking derivatives let $\mu = 0$. By doing so instead of the Eq.~(\ref{greenL}) we obtain
\begin{equation*}
(-1)^{-i/\gamma} \int_0^\infty  y^i p(x,t : y,t') dy = \frac{(-1)^{i/\gamma}}{\Gamma(i/\gamma)}\int_0^\infty e^{-x^{-\gamma } y Z(t,t')+ w(t,t') y^2}y^{i/\gamma-1}dy
\end{equation*}

The integral in the rhs part does not exist because $w(t,t') > 0$. This tells us that this particular case should be taken out of consideration since the model is not able to produce suitable values of volatility derivatives.

\section{Pricing European options on moment swaps}
European options on various moments of the volatility can be directly calculated if the density of the underlying process is known in closed form. Indeed, the price $C$ of, say the call option on $x^n_t$ by definition is
\begin{equation*}
C_n(x,T) = e^{-r T} \E\left[ \left(\frac{1}{T}\int_0^T x_t^n dt - K\right)^+ \ | \ x_0=x \right],
\end{equation*}
\noindent which can be rewritten as
\begin{equation*}
e^{r T} C_n(x,T) = \frac{1}{T}\int_0^T \left[\int_{K^{1/n}}^\infty (y^n - K) p(x,t,y) dy\right] dt
\end{equation*}
This double integral in general can not be computed in closed form, because of non-zero low limit of integration. Therefore, numerical integration is required. However, it can be done very efficiently, and also in parallel at any parallel architecture.

In the particular case of the model Eq.~(\ref{model2}) ($m = \gamma$), this could be done in closed form since the density function is just a delta function as it is presented in the Eq.~(\ref{trDelta}). Substituting this into the above formula and integrating we obtain
\begin{align} \label{optionsM2}
e^{r T} C_n(x,T) &= \alpha x^n -K = V_n - K \\
\alpha &= \frac{1}{T} \int_0^T e^{n \int_t^{t'} q(s) ds} \nonumber
\end{align}

It is important to underline that European options on volatility swaps can not be represented even in integral form under the proposed models and have to be computed numerically.

\section{Conclusions}
The main result of this paper is a set of new stochastic volatility models which consider the instantaneous variance (volatiltiy) process to be a mean-reverting CEV process. However, we do not fix the CEV power $\gamma$, but rather treat it as a model parameter to be determined by calibration. This approach meets the empirical observations presented in the literature and gives more flexibility when calibrating the model parameters to the market data on plain vanilla options and volatility derivatives.

The model preserves mean reversion and time dependence of the mean-reversion coefficient which also helps in calibrating term structure of variance swaps and other volatility derivatives. It also contains a non-linear drift term.

What distinguishes our proposed set of models from many possible models of this type
is that prices of the variance and moment swaps are obtained in closed form (by computation of one integral in time). They also give an approximate solution for the volatility swaps.
Also options on moments of the realized volatility are priced via computation of a double integral. This could be efficiently done using parallel calculations.

Despite our models do not allow a closed form solution for the characteristic or moment generations function, still they could be efficiently calibrated, e.g. to vanilla European option prices.
This could be done by using new efficient finite-difference methods or by using the mixing theorem. The latter means that conditional on the whole path of the instantaneous variance from $v_0$ to $v_T$ the option price $C_i$ then is given by the Black-Scholes formula with the efficient spot price $S^{e}$ and the efficient volatility $\sigma^{e}$ (see \cite{RomanoTouzi:97}). This method is very fast because its complexity is equivalent to one-dimensional MC.

\newpage

\begin{thebibliography}{}

\bibitem[Albanese et~al., 2001]{BSHyper2001}
Albanese, C., Campolieti, G., Carr, P., and Lipton, A. (2001).
\newblock {Black-Scholes} goes hypergeometric.
\newblock {\em Risk Magazine}, 14:99--103.

\bibitem[Bakshi et~al., 2004]{Bakshi2004}
Bakshi, G., Ju, N., and H.~Yang, H. (2004).
\newblock Estimation of continuous time models with an application to equity
  volatility.
\newblock Technical report, University of Maryland working paper.

\bibitem[Brockhaus and Long, 2000]{BrockhausLong2000}
Brockhaus, O. and Long, D. (2000).
\newblock Volatility swaps made simple.
\newblock {\em Risk}, pages 92--96.

\bibitem[Carr and Linetsky, 2006]{CarrLinetsky2006}
Carr, P. and Linetsky, V. (2006).
\newblock A jump to default extended {CEV} model: an application of {B}essel
  processes.
\newblock {\em Finance and Stochastics}, 10:303--330.

\bibitem[Carr and Sun, 2007]{CarrSun}
Carr, P. and Sun, J. (2007).
\newblock A new approach for option pricing under stochastic volatility.
\newblock {\em Review of Derivatives Research}, 10:87--250.

\bibitem[Chacko and Viceira, 1999]{ChackoViceira1999}
Chacko, G. and Viceira, L. (1999).
\newblock Spectral gmm estimation of continuous-time processes.
\newblock Technical report, Harvard University.

\bibitem[Craddock, 2009]{Craddock2009}
Craddock, M. (2009).
\newblock Fundamental solutions, transition densities and the integration of
  {Lie} symmetries.
\newblock {\em Journal of Differential Equations}, 246:2538--2560.

\bibitem[Craddock and Lennox, 2007]{CraddockLennox}
Craddock, M. and Lennox, K. (2007).
\newblock Lie group symmetries as integral transforms of fundamental solutions.
\newblock {\em J. Differential Equations}, 232:652--674.

\bibitem[Davydov and Linetsky, 2001]{DavidovLinetsky2001}
Davydov, D. and Linetsky, V. (2001).
\newblock Pricing and hedging path-dependent options under the {CEV} process.
\newblock {\em Management Science}, 47(7):949--965.

\bibitem[Gatheral, 2004]{Gatheral2005}
Gatheral, J. (2004).
\newblock A parsimonious arbitrage-free implied volatility parameterization
  with applicationto the valuation of volatility derivatives.
\newblock Global Derivatives And Risk Management.

\bibitem[Goldenberg, 1991]{Goldenberg:1991}
Goldenberg, D. (1991).
\newblock A unified method for pricing options on diffusion-processes.
\newblock {\em Journal of Financial Economics}, 29:3--34.

\bibitem[{Henry-Labord\'ere}, 2009]{Labordere2009}
{Henry-Labord\'ere}, P. (2009).
\newblock {\em Analysis, Geometry, and Modeling in Finance: Advanced Methods in
  Option Pricing}.
\newblock Chapman \& Hall/CRC Financial Mathematics Series.

\bibitem[Heston, 1993]{Heston:93}
Heston, S. (1993).
\newblock Closed-form solution for options with stochastic volatility, with
  applicationto bond and currency options.
\newblock {\em Review of Financial Studies}, 6(2):327--343.

\bibitem[{In't Hout} and Welfert, 2007]{HoutWelfert2007}
{In't Hout}, K.~J. and Welfert, B.~D. (2007).
\newblock Stability of {ADI} schemes applied to convection-diffusion
  equations with mixed derivative terms.
\newblock {\em Applied Numerical Mathematics}, 57:19--35.

\bibitem[Ishida and Engle, 2002]{IshidaEngle2002}
Ishida, I. and Engle, R. (2002).
\newblock Modelling variance of variance: The square root, the affine, and the
  {CEV GARCH} models.
\newblock Technical report, NYU.

\bibitem[Itkin and Carr, 2010]{ItkinCarr2010}
Itkin, A. and Carr, P. (2010).
\newblock Pricing swaps and options on quadratic variation under stochastic
  time change models - discrete observations case.
\newblock {\em Review Derivatives Research}, 13:141--176.

\bibitem[Javaheri, 2004]{Javaheri2004}
Javaheri, A. (2004).
\newblock {\em The volatility process: A study of stock market dynamics via
  parametric stochastic volatility models and a comparison to the information
  embedded in option prices}.
\newblock PhD thesis.

\bibitem[Jones, 2003]{Jones2003}
Jones, C. (2003).
\newblock The dynamics of stochastic volatility: evidence from underlying and
  options markets.
\newblock {\em Journal of Econometrics}, 116:118--224.

\bibitem[Lavoie et~al., 1976]{FracDer1976}
Lavoie, J., Osler, T., and Tremblay, R. (1976).
\newblock Fractional derivatives and special functions.
\newblock {\em SIAM Review}, 18:240--268.

\bibitem[Lee, 2004]{Lee2004}
Lee, R. (2004).
\newblock The moment formula for implied volatility at extreme strikes.
\newblock {\em Mathematical Finance.}, 14(3):469--480.

\bibitem[Lee, 2008]{Lee2008}
Lee, R. (2008).
\newblock Gamma swaps.
\newblock Technical report, University of Chicago.

\bibitem[Lewis, 2000]{Lewis:2000}
Lewis, A.~L. (2000).
\newblock {\em Option Valuation under Stochastic Volatility}.
\newblock Finance Press, Newport Beach, California, USA.

\bibitem[Lin, 2007]{Lin2007}
Lin, Y.-N. (2007).
\newblock Pricing VIX futures: Evidence from integrated physical and
  risk-neutral probability measures.
\newblock {\em The Journal of Futures Markets}, 27:1175--1217.

\bibitem[Olver, 1993]{Olver1993}
Olver, P. (1993).
\newblock {\em Applications of {Lie} Groups to Differential Equations}, volume
  107 of {\em Graduate Texts in Mathematics}.
\newblock Springer, New York, 2nd edition.

\bibitem[Revuz and Yor, 1999]{RevuzYor1999}
Revuz, D. and Yor, M. (1999).
\newblock {\em Continuous Martingales and Brownian Motion}.
\newblock Springer, Berlin, Germany, 3rd edition.

\bibitem[Romano and Touzi, 1997]{RomanoTouzi:97}
Romano, M. and Touzi, N. (1997).
\newblock Contingent claims and market completeness in a stochastic volatility
  model.
\newblock {\em Mathematical Finance}, 7(4):279--302.

\bibitem[Schoutens, 2005]{Schoutens2005}
Schoutens, W. (2005).
\newblock Moment swaps.
\newblock {\em Quantitative Finance}, 5(6):525--530.

\bibitem[Sepp, 2008]{Sepp2008}
Sepp, A. (2008).
\newblock Pricing options on realized variance in {Heston} model with jumps in
  returns and volatility.
\newblock {\em Journal of Computational Finance}, 11(4):33--70.

\bibitem[Swishchuk, 2004]{Swishchuk2004}
Swishchuk, A. (2004).
\newblock Modeling of variance and volatility swaps for financial markets with
  stochastic volatilities.
\newblock {\em WILMOTT Magazine}, 2:64--72.

\end{thebibliography}

\newcommand{\noopsort}[1]{} \newcommand{\printfirst}[2]{#1}
  \newcommand{\singleletter}[1]{#1} \newcommand{\switchargs}[2]{#2#1}

\end{document}